\documentstyle{article}
\title{On Complexity and Emergence}
\author{Russell K. Standish\\High Performance Computing Support
  Unit\\University of New South Wales\\http://parallel.hpc.unsw.edu.au/rks}
\begin{document}
\maketitle

\begin{abstract}
  Numerous definitions for {\em complexity} have been proposed over
  the last half century, with little consensus achieved on how to use
  the term. A definition of complexity is supplied here that is
  closely related to the Kolmogorov Complexity and Shannon Entropy
  measures widely used as complexity measures, yet addresses a number
  of concerns raised against these measures. However, the price of
  doing this is to introduce context dependence into the definition of
  complexity. It is argued that such context dependence is an inherent
  property of complexity, and related concepts such as entropy and
  emergence. Scientists are uncomfortable with such context
  dependence, which smacks of subjectivity, and this is perhaps the
  reason why little agreement has been found on the meaning of these terms.
\end{abstract}

\section{The problem of Complexity}

In the last 15 years, the study of {\em Complex Systems} has emerged
as a recognised field in its own right, although a good definition of
what a complex system actually is has eluded formulation. Attempts to
formalise the concept of {\em complexity} go back even further, to
Shannon's inception of {\em Information
  Theory}\cite{Shannon49}. A good survey of the tortuous path
the study of complexity has followed is provided in
Edmonds\cite{Edmonds99}. Of particular importance to this paper is the
concept of {\em Kolmogorov Complexity} (also known as {\em Algorithmic
  Information Complexity}) introduced independently by
Kolmogorov\cite{Kolmogorov65}, Chaitin\cite{Chaitin66} and
Solomonoff\cite{Solomonoff64}. Given a particular universal Turing
Machine\footnote{A Turing Machine is a formal model of a digital
  computer} (UTM) $U$, the Kolmogorov complexity of a string of
characters (ie a description) is the length of the shortest program
running on $U$ that generates the description.

There are two main problems with Kolmogorov complexity: 
\begin{enumerate}
\item The dependence on $U$, as there is no unique way of specifying
  this. Even though the Invariance theorem\cite[Thm 2.1.1]{Li-Vitanyi97}
  guarantees that any two UTMs U and V will agree on the complexity of
  a string $x$ up to a constant independent of $x$, for any
  descriptions $x$ and $y$, there will be two machines $U$ and $V$
  disagreeing on whether $x$ is more complex that $y$, or vice-versa.
\item Random sequences have maximum complexity, as
  by definition a random sequence can have no generating algorithm
  shorter than simply listing the sequence. As
  Gell-Mann\cite{Gell-Mann94} points out, this contradicts the notion
  that random sequences should contain no information.
\end{enumerate}

The first problem of what reference machine to choose is a symptom of
context dependence of complexity. Given a description $x$, any value
of complexity can be chosen for it by choosing an appropriate reference
machine. It would seem that complexity is in the eye of the
beholder\cite{Emmeche94}. Is complexity completely subjective? Is
everything lost? 

Rather than trying to hide this context dependence, I would prefer to
make it a feature. Instead of asserting complexity is a property of some
system, it is a property of descriptions (which may or may not be
about a system). There must also be an interpreter of these descriptions
that can answer the question of whether two descriptions are equivalent
or not. Consider Shakespeare's {\em Romeo and Juliet}. In Act II,ii,
line 58, Juliet says ``My ears have yet not drunk a hundred
words''. If we change the word ``drunk'' to ``heard'', your average
theatre goer will not spot the difference. Perhaps the only one to
notice would be a professional actor who has played the scene many
times. Therefore the different texts differing by the single word
``drunk/heard'' in this scene are considered equivalent by our
hypothetical theatre goer. There will be a whole {\em equivalence
  class} of texts that would be considered to be {\em Romeo and
  Juliet} by our theatre goer.

Once the set of all possible descriptions are given (strings of letters
on a page, base pairs in a genome or bits on a computer harddisk for
example), and an equivalence class between descriptions given, then
one can apply the Shannon entropy formula to determine the complexity
of that description, under that interpretation:
\begin{equation}\label{complexity}
C(x) = \lim_{\ell\rightarrow\infty} \ell\log_2 N - \log_2 \omega(\ell,x)
\end{equation}
where $C(x)$ is the complexity (measured in bits), $\ell(x)$ the length of
the description, $N$ the size of the alphabet used to encode the
description and $\omega(\ell,x)$ the size of the class of all descriptions
of length less than $\ell$ equivalent to $x$. We assume that the
interpreter of a description is able to determine where a description
finishes, so that a description $y$ of length $\ell(y)$ is equivalent
to all $N^{\ell-\ell(y)}$ length $\ell$ descriptions having $y$ as a prefix.

If we choose our description set to be all bitstrings, and our
equivalence class to be all bitstrings that produce the same output
when executed by a universal Turing machine $U$, then 
\begin{equation}
\omega(\ell,x) = \sum_{p:U(p)=U(x), \ell(p)\leq\ell} 2^{\ell-\ell(p)},
\end{equation}
where $\ell(p)$ is the length of program $p$. As
$\ell\rightarrow\infty$, this distribution (when normalised) is known
as the {\em universal a priori probability}\cite[Def
4.3.3]{Li-Vitanyi97}. By the {\em coding theorem}, the complexity
defined by equation (\ref{complexity}) is equal to the Kolmogorov
complexity up to a constant independent of $x$.

This perspective helps us understand the problem of random strings
having maximal complexity. In an equivalence class generated by a
human observer, one random string is pretty much the same as any
other. Therefore the $\omega$ term of a completely random string will
large, probably of comparable size to $N^\ell$. Therefore the
complexity of a random string, as interpreted by a human observer is
low, exactly the property required of Gell-Mann's {\em effective
  complexity}.

Whilst context dependence would appear to open up the curse of
subjectivity, it needn't necessarily do so. In many situations, the
equivalence relation is well defined. For example the notion of
species in biology is reasonably well defined (although disagreement
exists in a number of cases). This could, in principle, along with a
detailed knowledge of the genetic code, be used to estimate the
complexity of different species. This principle has been used in a
number of artificial life studies\cite{Adami99a,Standish99a} for
studying the evolution of complexity. 

\section{Emergence}

Emergence is that other area of complex systems study that has
experienced controversy and confusion. Its importance stems from the
belief that emergence is the key ingredient that makes complex systems
complex. Putting things colloquially, emergence is the concept of some
new phenomenon arising in a system that wasn't in the system's
specification to start with. There is some considerable debate as to
how this happens, or whether emergence can truly happen within a
formal system such as an agent-based
model\cite{Rosen99,Funtowicz-Ravetz94}.

Let me illustrate this debate with the example of gliders in {\em The
  Game of Life}. I would contend that this phenomenon is emergent, in
  the sense that the glider concept is not contained within the
  cellular automaton implementation language --- namely states
  and neigbourhoods. This puts me at odds with Rosen, who would argue
  that gliders are but complicated combinations of simple machines
  (the cellular automata), not examples of {\em
  complexity}\footnote{Rosen uses the word complexity as a quality,
  rather like  emergence is used in this paper, as opposed to a
  quantity.} 

What is my definition of emergence then? To set the scene, let me
introduce two descriptions of a system, called the {\em
  microdescription} and {\em macrodescription}, each coded in their
own language. Ronald et al., use the term ${\cal L}_1$ and ${\cal
  L}_2$ to refer to the micro- and macrolanguage
respectively\cite{Ronald-etal99}. They call the microlanguage the
``language of design'' in view of artificial life
applications. However, the microdescription may equally well be our
best description of what happens at the most fundamental level. To
make it clear, emergence is not due to the failure of the
microdescription as a modeling effort, since the emergent property
should still appear as the result of a computer simulation constructed
using the microdescription.

We also assume that the macrodescription is a {\em good theory}. There
is, in general, a trade-off between the predictive power of a theory,
and its explanatory power. Of course, a theory may be neither
predictive nor explanatory, but in this case the theory is not so
good, and would be rejected in favour of one that is better. One can,
of course, produce examples of emergent concepts based on a poor
macrodescription, but in this case the correspondence with the real or
simulated system would be lacking, and the concept would not be an
observed phenomenon.

An emergent phenomenon is simply one that is described by atomic
concepts available in the macrolanguage, but cannot be so described in
the microlanguage. In the case of the glider in The Game of Life, any
attempt at describing a glider would involve the CA transition table
(naturally), but also the specific pattern of cell states that make up
the glider. But which pattern? A glider can appear at any location
within the CA, and may have one of four possible orientations. The
description cannot represent the fact that two gliders separated
diagonally by 1 cell in along each axis with the same orientation are
temporally related. A glider, as an object-in-itself, is a pure
macrodescription object.

Ronald et al. focus on the element of {\em surprise} as a test of
emergence. In this, they are trying to capture emergence as some kind
of dissonance between the micro and macro languages. To be fair to
their work, they claim only a test for emergence, not a definition,
along the lines of the more famous Turing test. However, the surprise
factor really only works when the macrolanguage is enlarged (by the
emergent concept) in order to make the macrodescription a better model
of the system. Once the emergent property has been recognised by the
observer, the property is no longer surprising. The definition of
emergence given here works, regardless of whether the observer is
still surprised or not.

Of considerable interest is, given a system specified in its
microlanguage, does it have emergent properties? There is no general
procedure for answering this question. One has to construct a
macrodescription of the system. If this macrodescription contains
atomic concepts that are not simple compounds of microconcepts, then
one has emergent properties. Is there a best macrodescription for any
given system? This question is outside of the scope of this paper, and
needs to be answered by theories of how scientific theories are
developed. However, in general, it seems unlikely that there would be
a ``best description'', as it depends on the motives of the person
using the description. For example, I have already alluded to the
tension between predictive power and explanatory power.

\section{Entropy as a case study of emergence}

{\em Thermodynamics} provides an excellent case study in emergence,
as it is well understood theory, and also illustrates the link between
emergence and complexity.

Thermodynamics is a macroscopic description of material systems,
expressed in terms of concepts such as temperature, pressure and
entropy. It is related to the microscopic description of molecular
dynamics via the reductionist theory of {\em statistical mechanics}.

Within thermodynamics, entropy is defined by differences along a
reversible path:
\begin{displaymath}
\Delta S = \Delta Q/T
\end{displaymath}
Within the framework of thermodynamics, this quantity is objectively
defined, up to an additive constant (usually assumed to be such that
entropy vanishes at absolute zero).

Within Statistical Mechanics, entropy in the microcanonical ensemble
is given by the Boltzmann formula:
\begin{equation}\label{entropy}
S = k_B \ln W,
\end{equation}
where $k_B$ is the Boltzmann constant (giving entropy in units of
Joules per Kelvin), and $W$ is the number of microstates accessible to
the system for a particular macrostate.  This formula is very similar
to the information-based complexity formula (\ref{complexity}), and
this led Jaynes\cite{Jaynes65} to remark that entropy ``measures our
degree of ignorance as to the unknown microstate''. Denbigh and
Denbigh\cite{Denbigh-Denbigh87} are at pains to point out that entropy
is fully objective, as it only depends on the macroscopic quantities
chosen to define the macroscopic state (the temperature, pressure and
so on), and not dependent on any extraneous observer related property.
Provided two observers choose the same microscopic and macroscopic
description of the system, they will agree on the value of entropy. By
my previous definition of emergence, entropy is an emergent quality of
the system. This example illustrates how context dependent emergent
properties can be fully objective.

With entropy defined by (\ref{entropy}), the well known H-theorem
holds. As a consequence, the macroscopic (thermodynamic) description
is time irreversible, whereas the microscopic description is
reversible. Time irreversibility is likewise an emergent property of
this system.

The very clear relation between the Boltzmann-Gibbs entropy
(\ref{entropy}), and complexity (\ref{complexity}) indicates that
complexity is itself an emergent concept. If the microscopic language
and macroscopic language were identical, corresponding to a situation
of no emergence, complexity of descriptions degenerates to the trivial
measure of description length.

\bibliographystyle{plain}
\bibliography{rus}
\end{document}